\documentclass[prl,10pt,floatfix,amsmath,twocolumn,showpacs,superscriptaddress]{revtex4-1}

\usepackage{graphicx}  
\usepackage{dcolumn}   
\usepackage{bm}        
\usepackage{amssymb}   
\usepackage{amsmath}
\usepackage{amssymb}
\usepackage{graphicx}
\usepackage{amsfonts}%
\usepackage{dcolumn}
\usepackage{bm}
\usepackage{hyperref}
\usepackage{nccmath}
\usepackage{lipsum}
\usepackage{multirow} 
\usepackage{color,xcolor}
		{\end{pmatrix}\end{medsize}}%

\begin{document}
\title{ Topological Luttinger semimetallic phase accompanied with surface states realized in silicon }
\author{Ying Li}
\affiliation{Wuhan National High Magnetic Field Center $\&$ School of Physics,
Huazhong University of Science and Technology, Wuhan 430074, China}
\author{Chi-Ho Cheung}
\affiliation{Wuhan National High Magnetic Field Center $\&$ School of Physics,
Huazhong University of Science and Technology, Wuhan 430074, China}
\author{Gang Xu}\email{gangxu@hust.edu.cn}
\affiliation{Wuhan National High Magnetic Field Center $\&$ School of Physics,
Huazhong University of Science and Technology, Wuhan 430074, China}
\date{\today}


\begin{abstract}
By means of systematically first-principles calculations and model analysis, a complete phase diagram of the body-centered silicon(BC8-Si) via lattice constant $a$ and internal atomic coordinate $x$ is explored, which demonstrates that BC8-Si is a topological Luttinger semimetal(LSM) accompanied with topologically nontrivial surface states, and the electronic properties of BC8-Si can be further tuned to a normal insulator or topological Dirac semimetal by very tiny changing of $a$ and $x$. These results successfully explain the contradictory transport reports of BC8-Si. More importantly, the topological surface states in the LSM phase fill in the gap between the topological matters and silicon, which provide an opportunity to integrate the topological quantum devices and silicon chips together.

\end{abstract}

\pacs{ }

\maketitle

\textit{Introduction}---Silicon is the most important material for electronic~\cite{4785577, 4785580, kittel1996introduction} and photovoltaic industry ~\cite{carlson1976amorphous, 1982ph...book.....G, green2020solar, PhysRevB.86.121204} due to its excellent electronic properties and mature technology.
Especially, almost 90\% of electronic chips are equipped based on the diamond cubic silicon (DC-Si). Nonetheless, the miniaturization of Si-based chips is facing the end of Moore's Law due to the limitation of quantum effects~\cite{khan2018science}. Searching and devising the next-generation of electronic devices are the most urgent and challenging task~\cite{radisavljevic2011single,hsieh2008topological}.


In the past decades, topological matters with nontrivial boundary modes have attracted intensive attentions due to their novel properties~\cite{kane2005quantum, qi2011topological, xu2011chern, hua2018dirac,zou2019study}, such as backscattering suppression~\cite{buttiker1988absence, yan2012topological, xu2015quantum, xu2015intrinsic}, spin-momentum locking~\cite{hsieh2009tunable, gotlieb2018revealing, zhang2013spin, nie2017topological} and non-abelian braiding~\cite{nayak2008non, wu2020pursuit, xu2016topological}, which are expected to be a significant platform for the next-generation electronic and spintronic devices~\cite{qi2009inducing, garate2010inverse}.
In order to integrate the topological quantum devices and silicon chips together, it is highly desirable to realize the topological boundary modes in silicon. However, these two fields have no overlap until now, because DC-Si is well known as semiconductor without band inversion, which makes it impossible to hold the topological boundary modes.


Fortunately, silicon has more than 13 allotropes~\cite{ackland2019high, haberl2016pathways}. Among them, a body-centered cubic structure, named as BC8-Si, was reported to be stabilized under ambient conditions~\cite{wentorf1963two, crain1994reversible, piltz1995structure},
but its electronic properties are under debate~\cite{malone2008ab, Zhang2017BC8, besson1987electrical, wosylus2009crystal, pfrommer1997ab}.
While previous experiments and calculations suggest that BC8-Si is a semimetal with band overlap~\cite{besson1987electrical, wosylus2009crystal, pfrommer1997ab, malone2008ab}, a contrary literature reports that it is a narrow band gap semiconductor recently~\cite{Zhang2017BC8}. In particular, the topological properties of BC8-Si have never been studied yet.
In this paper, by means of the first-principles calculations, we investigate the electronic and topological properties of BC8-Si systematically, and demonstrate
that it is a topological Luttinger semimetal (LSM) with
band inversion, on the surface of which, topological surface states can be stabilized.
Moreover, our numerical results indicate that the electronic and topological properties of BC8-Si are sensitive to the lattice constant $a$ and internal atomic coordinate $x$.
A complete phase diagram via $a$ and $x$ is explored, which demonstrates that the topological LSM phase of BC8-Si can be tuned to a normal insulator (NI) without band inversion or a topological Dirac semimetal (DSM) by very tiny changing of $a$ and $x$. Such changing can be achieved by varying the applied pressure during crystal synthesis~\cite{crain1994tetrahedral, hu1986crystal, kasper1964crystal, kurakevych2016synthesis}. Our results successfully explain the controversial reports on electronic properties of BC8-Si. More importantly, the topological properties of BC8-Si fill in the gap between the topological matters and silicon, which provides an opportunity to integrate the next-generation electronic quantum devices and silicon chips together.

\begin{figure}
\includegraphics[width=\columnwidth]{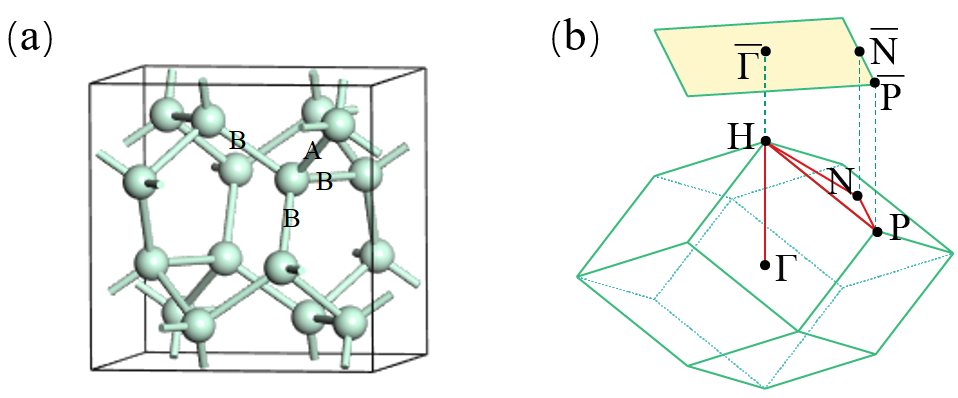}
\caption{(a) Unit cell of the body-centered cubic silicon, BC8-Si,
with space group $Ia-3$ (No.~206),
where two types of Si-Si bond length A=2.305~\AA, B = 2.391~\AA~are labeled.
(b) BZ  of the primitive cell with high-symmetry path and its projection on the (001) surface of the unit cell.
\label{fig:fig1}}
\end{figure}

\textit{Crystal structure and methodology} ---As shown in Fig.~\ref{fig:fig1}(a),
BC8-Si adopts the body-centered cubic lattice with space group $Ia$-3 (No.206), where the lattice constant $a$ equals 6.636~\AA~and Si atoms are located at $16c$ Wyckoff position with coordinate $x=0.1003$~\cite{kasper1964crystal}.
These most reported experimental crystal parameters are used in our calculations,
otherwise they will be explicitly pointed out. Compared with DC-Si,
Si atoms form a slightly distorted tetrahedral structure with two types Si-Si
distance $A = 2.305$~\AA~and $B = 2.391$~\AA~in BC8-Si.
The first Brillouin zone (BZ) of the primitive cell and its projection on the (001) surface of the unit cell
are displayed in Fig.~\ref{fig:fig1}(b).
Our first-principles calculations are performed by the Vienna ab initio simulation package~\cite{kresse1996efficient, kresse1996efficiency} with the projected augmented wave method ~\cite{blochl1994projector}. The energy cutoff is set as 400 eV, and $7\times7\times7$ k-meshes are adopted. Perdew-Burke-Ernzerhof type of the exchange-correlation potential~\cite{perdew1996generalized}, and Heyd-Scuseria-Ernzerhof (HSE06) hybrid functional~\cite{heyd2003hybrid} with Hartee-Fock exchange factor 0.35 are used in all calculations to obtain the accurate electronic structures. Spin-orbit coupling (SOC) interaction is considered consistently. Symmetry preserving Wannier functions are constructed by the Wannier90 package~\cite{mostofi2014updated}. The surface states are calculated by iterative Green's function method as implemented in the WannierTools package~\cite{wu2018wanniertools}.


\begin{figure}
\includegraphics[width=\columnwidth]{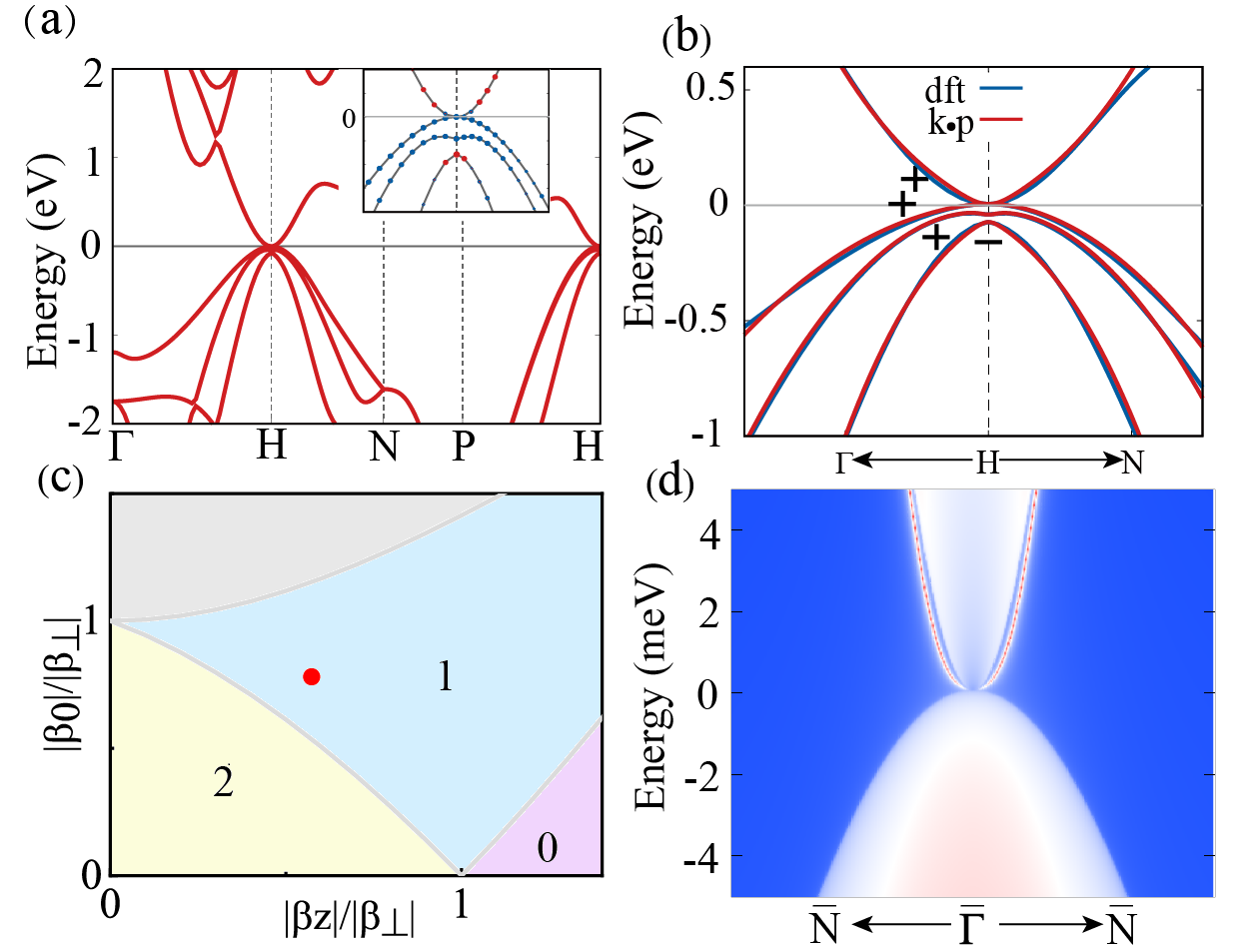}
\caption{(a) The calculated band structures of BC8-Si,
red and blue circles indicate the projection to the bonding states and antibonding states, respectively.
(b) The fitted band structures by the eight-band $k \cdot p$ model and first-principles calculations.
(c) The topological phase diagram of the 2D nodal semimetal in the ($\beta_0, \beta_z$) plane, where the parameters of BC8-Si's reduction is marked by the red circle.
(d) The calculated LDOS illustrates the topological surface mode on the $(001)$ surface of the unit cell.
\label{fig:fig2}}
\end{figure}

\textit{Luttinger semimetal} --- The electronic configuration of Si atom is $3s^23p^2$. Because each Si atom
is tetrahedrally connected with other four Si atoms,
the main chemical bonding in BC8-Si is $sp^3$ hybridization~\cite{haerle2001sp},
which is similar to the chemical bonding in DC-Si~\cite{bazant1997environment}.
As a result, the calculated band structures in Fig.~\ref{fig:fig2}(a)
show that the $sp^3$ bonding states are almost fully occupied and contribute to the valence bands,
while the $sp^3$ antibonding states are almost empty and form conduction bands.
However, different from semiconductor DC-Si, one antibonding
state in BC8-Si is lower than its bonding states
at $H$ point as shown in the inset of Fig.~\ref{fig:fig2}(a) and Fig.~\ref{fig:fig2}(b),
which leads to a band inversion in BC8-Si and makes it a semimetal.
Due to the presence of inversion symmetry ($I$)
and time reversal symmetry ($\mathcal{T}$),
each band in BC8-Si is doubly degenerate at every momentum $k$,
which means that the band touching at Fermi level ($E_F$) is fourfold degeneracy.
Furthermore, the dispersions around the fourfold degenerate node are always parabolic as shown in Fig.~\ref{fig:fig2}(b). Accordingly,
BC8-Si is a LSM as reported by previous experimental
and theoretical studies~\cite{besson1987electrical,wosylus2009crystal, pfrommer1997ab, malone2008ab}.

\textit{Model analysis} --- To understand the low energy band structures in BC8-Si,
we construct a $k \cdot p$ model at $H$ point, where the vector group is $T_h$ with generators $C_{2x}$, $C_3^{111}$ and $I$.
By analysing the low energy bands' irreducible representations(IRRs) at $H$ point,
it shows that the antibonding state is equivalent to an effective s-orbital with odd parity,
while the three bonding states equal to three effective p-orbitals with even parity.
Under the basis of $| p_x\rangle^+$, $|p_y\rangle^+$, $| p_z\rangle^+$ and $|s\rangle^-$
with the superscript +/- denoting parity, the generators of
$T_h$ can be expressed by the matrix form as shown in Eq.~\ref{eq:1},
and the Hamiltonian without SOC is constructed as Eq.~\ref{eq:H_{non-SOC}}((see details
in appendix A).
In Table.~\ref{tab:1}, the parameters of the effective model are obtained by fitting with first-principles calculations.
The fitted band structures are plotted in Fig.~\ref{fig:non-SOCfit},
which give rise to a triply degenerate node formed
by the bonding orbitals at $H$ point due to the protection of $C_3^{111}$.

By considering SOC, the Hamiltonian under
the spinful basis ($|p_x\rangle^+$, $|p_y\rangle^+$, $|p_z\rangle^+$, $|s\rangle^-$) $\otimes$
($|\uparrow\rangle$  $|\downarrow\rangle$) is given in Eq.~\ref{eq:Hsoc},
where the $H_{soc}$ is shown in Eq.~\ref{eq:p_{soc}} with atomic SOC strength parameter $\lambda$.
With $\lambda=0.0152$~eV,
the energy eigenvalues of the eight-band model well reproduce the band structures
from first-principles calculations as shown in Fig.~\ref{fig:fig2}(b) and Fig.~\ref{fig:socfit}.
We further transform Eq.~\ref{eq:Hsoc} to the SOC representation $|J, J_z\rangle$ as described
in appendix B, where $J = L+S$ is the total angular momentum,
and $J_z$ is its projections on the z-axis.
Under the basis of $| \frac{3}{2},  \frac{1}{2}\rangle^{+} $, $| \frac{3}{2},  \frac{3}{2}\rangle^{+} $, $| \frac{3}{2},  -\frac{3}{2}\rangle^{+} $, $| \frac{3}{2}, - \frac{1}{2}\rangle^{+} $, $| \frac{1}{2},  \frac{1}{2}\rangle^{+} $, $| \frac{1}{2},  -\frac{1}{2}\rangle^{+} $, $| \frac{1}{2},  \frac{1}{2}\rangle^{-} $ and $| \frac{1}{2}, - \frac{1}{2}\rangle^{-} $,
the new Hamiltonian is written in Eq.~\ref{eq:H_{basis8}},
which explicitly shows that the $|J=\frac{1}{2}\rangle^+$ doublet are pushed down 2$\lambda$
while $|J=\frac{3}{2}\rangle^+$ quartet are pushed up $\lambda$ at $H$ point by SOC interaction.
As a result, the energy difference between $|J=\frac{1}{2}\rangle^-$ and $|J= \frac{3}{2}\rangle^+ $
at $H$ point can be described as $\Delta=E_{ab}-E_b-\lambda$.
According to the fitted parameters list in Table.~\ref{tab:1}, we get $\Delta$= -78~meV $< 0$,
which means that the band inversion is well reproduced by our model and parameters. One important consequence
of the band inversion is that the fourfold degenerate node formed by the $|J= \frac{3}{2}\rangle^+$ states
is  leaved at the $E_F$ exactly. Since $k^2$ is the leading order
in the $|J=\frac{3}{2}\rangle^+$ subspace, the band dispersions around the fourfold
degenerate node are always parabolic.
These results theoretically clarify that 
BC8-Si is a LSM with band inversion.

\begin{table}[htbp]\footnotesize
\centering
\setlength{\tabcolsep}{0.3mm}
\caption{Fitted parameters of the eight-band $k \cdot p$ Hamiltonian. $E_b$ and $E_{ab}$ are the on-site energies of bonding and antibonding states
at $H$ point. $t_1$, $t_2$, $t_3$ and $t_4$ are the mass terms
in bonding and antibonding subspace, respectively.
$t_5$ is the coupling between the bonding and antibonding subspaces,
while $t_6$ is the coupling within the bonding subspace.}
\newcommand{\tabincell}[2]{\begin{tabular}{@{}#1@{}}#2\end{tabular}}
\begin{tabular}{cccccccccc}
\hline
$t_1$ & $t_2$ & $t_3$ &$t_4$ & $t_5$ &$t_6$ & $E_b$ & $E_{ab}$ &$\lambda$ \\
eV$\cdot$\AA$^2$&eV$\cdot$\AA$^2$&eV$\cdot$\AA$^2$&eV$\cdot$\AA$^2$&eV$\cdot$\AA&eV$\cdot$\AA$^2$&eV&eV&eV\\
 \hline
$-4.638$ & $-8.636$ & $-20.098$ & $11.662$ & $-2.413$ & $-8.5$ & $0.0$ & $-0.063$ & $0.015$ \\ \hline
\end{tabular}
\label{tab:1}
\end{table}

\textit{Topological characters} ---
Maxim Kharitonov et al.~have confirmed that the four-band 3-dimentional LSM is topological and can exhibit surface states,
if its 2-dimentional(2D) reductions to some planes in momentum space passing the
quadratic node are topologically nontrivial~\cite{kharitonov2017universality}.
The topological properties of the reduced 2D nodal semimetal
is determined by the phase diagram in ($\frac{|\beta_0|}{|\beta_\perp|}, \frac{|\beta_z|}{|\beta_\perp|}$)
parameter plane, as shown in Fig.~\ref{fig:fig2}c.
Here, $\beta_\perp$ is the coefficient of the pauli matrix $\sigma_{x,y}$
characterizing a chiral symmetric of 2D LSM, and $\beta_0,~\beta_z$
are the coefficients of $\sigma_0$ and $\sigma_z$ describing the breaking of the chiral symmetry.
In Fig.~\ref{fig:fig2}c, the gray region $\lvert\beta_0\rvert>\sqrt{\lvert\beta_{\perp}\rvert^2+\beta_z^2}$
means that the system is no longer a semimetal,
the pink region labeled by 0 is trivial semimetal,
the blue and yellow region labeled by topological number 1 and 2
are the non-trivial LSM accompanied of one and two edge states, respectively.

To explore the topological properties of BC8-Si,
the eight-band model in Eq.~\ref{eq:H_{basis8}} is downfolded to four-band Hamiltonian
by perturbation theory~\cite{lowdin1951note},
the details of downfolding are shown in appendix C.
Under the basis of $| \frac{3}{2},  \frac{1}{2}\rangle^{+} $, $| \frac{3}{2},  \frac{3}{2}\rangle^{+} $, $| \frac{3}{2},  -\frac{3}{2}\rangle^{+} $
and $| \frac{3}{2}, - \frac{1}{2}\rangle^{+} $,
a simple Hamiltonian describing the low energy physics of LSM in BC-8 Si
is written as Eq.~\ref{eq:4} up to its leading term of $k$:
\begin{flalign}
\begin{split}
&H_{4\times4} = H_0 + H^\prime,\\
&H_0=\begin{bmatrix}
M_0 (k) & -\frac{t_6 k_+ kz}{\sqrt{3}} & V_1 (k)& 0\\
-\frac{t_6 k_- kz}{\sqrt{3}} & M_1 (k) & 0 & V_1 (k) \\
V_1^{*}(k) & 0 & M_1 (k) & \frac{t_6 k_+ kz}{\sqrt{3}} \\
0	 & V_1^{*}(k) & \frac{t_6 k_- kz}{\sqrt{3}} & M_0 (k)\end{bmatrix}\!,\\
&H^\prime=\frac{t_5^2}{-\Delta}\begin{bmatrix}
\frac{1}{6}k_+k_- +\frac{2k_z^2}{3} & -\frac{k_+k_z}{\sqrt{3}} & -\frac{k_-^2}{2\sqrt{3}} & 0\\
-\frac{k_- kz}{\sqrt{3}} & \frac{k_+k_-}{2} & 0 & -\frac{k_-^2}{2\sqrt{3}} \\
-\frac{k_+^2}{2\sqrt{3}}	 & 0 & \frac{k_-k_+}{2}&  \frac{k_+k_z}{\sqrt{3}}\\
0	 & -\frac{k_+^2}{2\sqrt{3}}  & \frac{k_-k_z}{\sqrt{3}} & \frac{1}{6}(k_-k_+) +\frac{2k_z^2}{3}&
\end{bmatrix}
\end{split}
\label{eq:4}
\end{flalign}where $H_0$ comes from the upper left $4 \times 4$ block ($|J=\frac{3}{2}\rangle^+$ subspace) of the eight-band model in Eq.~\ref{eq:H_{basis8}},
with $M_0(k)=E_b+\lambda + \frac{1}{6}[(t_1 +4t_2 +t_3 ) k^2_{x} + (4t_1 +t_2 +t_3 )k^2_{y} + (t_1 +t_2 +4t_3 )k^2_{z}]$, $ M_1 (k)=E_b+\lambda + \frac{1}{2}[t_1(k^2_{x} +k^2_{z} ) + t_2(k^2_{y} +k^2_{z} ) + t_3(k^2_{x} +k^2_{y})]$,
$V_1 (k)=\frac{1}{2\sqrt{3}}[k^{2}_{z} (-t_1 + t_2 ) + k^{2}_{y} (-t_2 + t_3 ) + k^{2}_{x} (t_1 - t_3 ) + 2it_6 k_x k_y]$,
$k^{2} =k_{x}^{2} +k_{y}^{2} +k_{z}^{2} $, $k_\pm = k_x \pm ik_y $.
$H^\prime$ is the influence part from $|J_z=\pm\frac{1}{2}\rangle^-$ states.
By assuming $k_z=0$, we reduce Eq.~\ref{eq:4} to a  $4\times4$ 2D model passing
the quadratic node.
Moreover, due to the $\mathcal{T}$ symmetry,
the $4\times4$ 2D model can be further decoupled into two $2\times2$ 2D models.
One is the under the basis of $ | \frac{3}{2},  \frac{3}{2}\rangle^{+} $, $| \frac{3}{2}, - \frac{1}{2}\rangle^{+}$
as written in Eq.~\ref{eq:H_N3}, and the other is its conjugate under basis
of $| \frac{3}{2},  -\frac{3}{2}\rangle^{+} $, $| \frac{3}{2},  \frac{1}{2}\rangle^{+}$.
Thus, we can just study the coefficients of Eq.~\ref{eq:H_N3} separately to
determine the topological properties of the 2D quadratic model
as shown in appendix D.
With parameters given in Table~\ref{tab:1},
the calculated $\frac{|\beta_0|}{|\beta_\perp|}, \frac{|\beta_z|}{|\beta_\perp|}$ are 0.7804 and 0.5733,
which corresponds to a phase point as marked by a red circle in Fig.~\ref{fig:fig2}(c) that belongs
to the nontrivial region with topological number 1.
Our numerical results strongly imply that BC8-Si is a topological LSM,
and one surface state is anticipated on the surface.
To prove that, the maximally localized Wannier functions for antibonding
and bonding orbitals are constructed.
The local density of states (LDOS) on the (001) surface are calculated
based on the maximally localized Wannier functions by using Green's function method.
As shown in Fig.~\ref{fig:fig2}(d),
one topological surface state at
$\bar{\Gamma}$ is presented clearly,
which confirms the topologically nontrivial characters of BC8-Si.
We note that the topological surface states exhibit a parabolic dispersion as shown in Fig.~\ref{fig:fig2}(d), which is different from the surface states in other topological materials such as topological insulator or topological DSM\cite{zhang2009topological, wang2012dirac}. Further angle-resolved photoemission spectroscopy experiment is highly desirable to verify such topological surface states in BC8-Si.

\begin{figure}
	\includegraphics[width=\columnwidth]{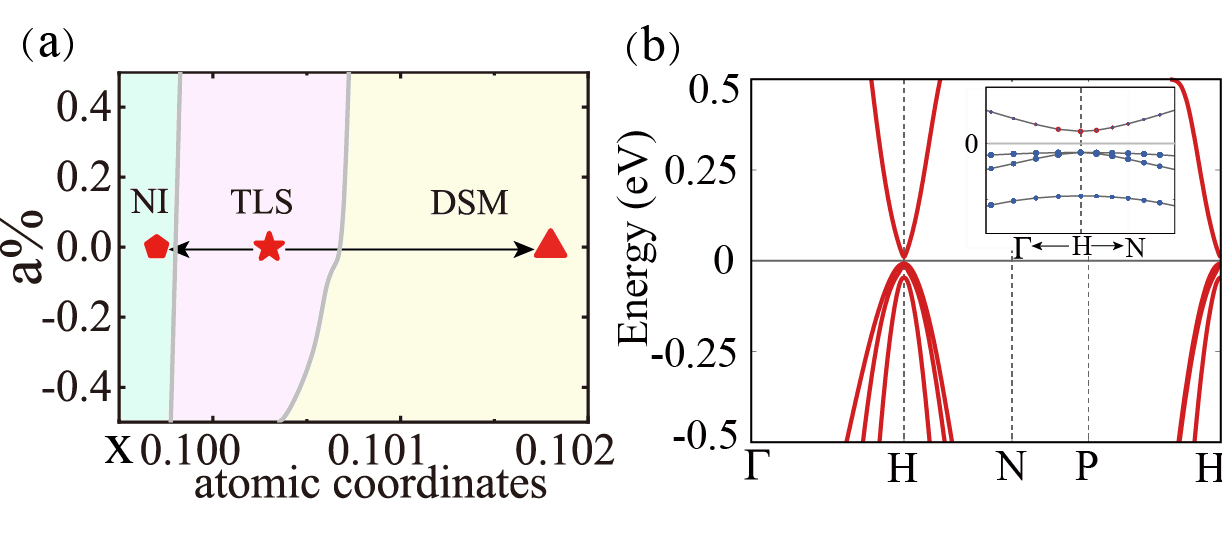}
	\caption{(a) Phase diagram of BC8-Si under strain in the range of (-0.5$\%$, +0.5$\%$).
    (b) The band structures of NI, the inset is the projected band structures around $H$ point, red and blue circles indicate the projection to the bonding states and anti-bonding states, respectively.
    \label{fig:fig3}}
\end{figure}

\textit{Phase diagram} ---
Since it has been reported that the
electronic property of BC8-Si is sensitive to lattice constant a and internal atomic coordinate $x$~\cite{Zhang2017BC8, malone2008ab}, we
further investigate their impacts on the band structures
and topological properties by first principles calculations.
The calculated phase diagram with various $a$ and $x$ consists of three subregions,
NI, topological LSM and DSM, labeled by different colors in Fig.~\ref{fig:fig3}(a).
Our results demonstrate that the electronic properties of BC8-Si
are more sensitive to the atomic coordinate $x$.
Keeping $a = 6.636$ \AA~as a constant ($a\%$ =0.0 in Fig.~\ref{fig:fig3}(a)),
it shows that: (1) BC8-Si is a NI without band inversion when $x$ is less than 0.0998;
(2) it falls into the topological LSM phase for the medium value
of the reported results $0.1008 > x > 0.0998$~\cite{crain1994tetrahedral, hu1986crystal, kasper1964crystal},
where the parameters used in Fig.~\ref{fig:fig2} are marked by a red star;
(3) BC8-Si becomes topological DSM consisting both the Dirac point and the quadratic node, when $x$ exceeds 0.1008.
In general, pressure (strain) will shorten (enlarge) the distance between the Si atoms
by increasing (decreasing) $x$ and compressing (elongating) $a$~\cite{wang2014direct},
which indicates the various electronic states of BC8-Si illustrated in  Fig.~\ref{fig:fig3}(a),
as discussed in $\alpha$-Sn~\cite{zhang2018engineering} and $\mathrm{Cu_2Se}$~\cite{zhu2018quadratic}.

\textit{Normal insulator} ---Actually, different structural parameters,
and even different electronic states can be accessed
by varying the synthesis conditions~\cite{wang2014direct,crain1994tetrahedral, hu1986crystal, kasper1964crystal}.
Previous optical
spectroscopy and electrical conductivity measurements~\cite{Zhang2017BC8}
indeed show that BC8-Si fall into the semiconducting phase with
an ultra narrow direct band gap,
which corresponds well to the NI phase in Fig.~\ref{fig:fig3}(a).
In Fig.~\ref{fig:fig3}(b),
by using $a = 6.636$~\AA~and $x = 0.0997$,
we plot the calculated band structures as a representation of the NI state,
which gives rise to a direct band gap of 5 $meV$ at $H$ point.
Different from the topological LSM phase,
the energy difference $\Delta$ is positive in the NI phase,
which is clearly illustrated by the orbital projections shown
in the inset of Fig.~\ref{fig:fig3}(b).
Finally, we would like to emphasize that the coordinate
$x$ only changes about  0.6$\%$ from the topological LSM in
Fig.~\ref{fig:fig2} to this NI phase, which is totally feasible by changing the applied pressure during crystal synthesis.


\begin{figure}
	\includegraphics[width=\columnwidth]{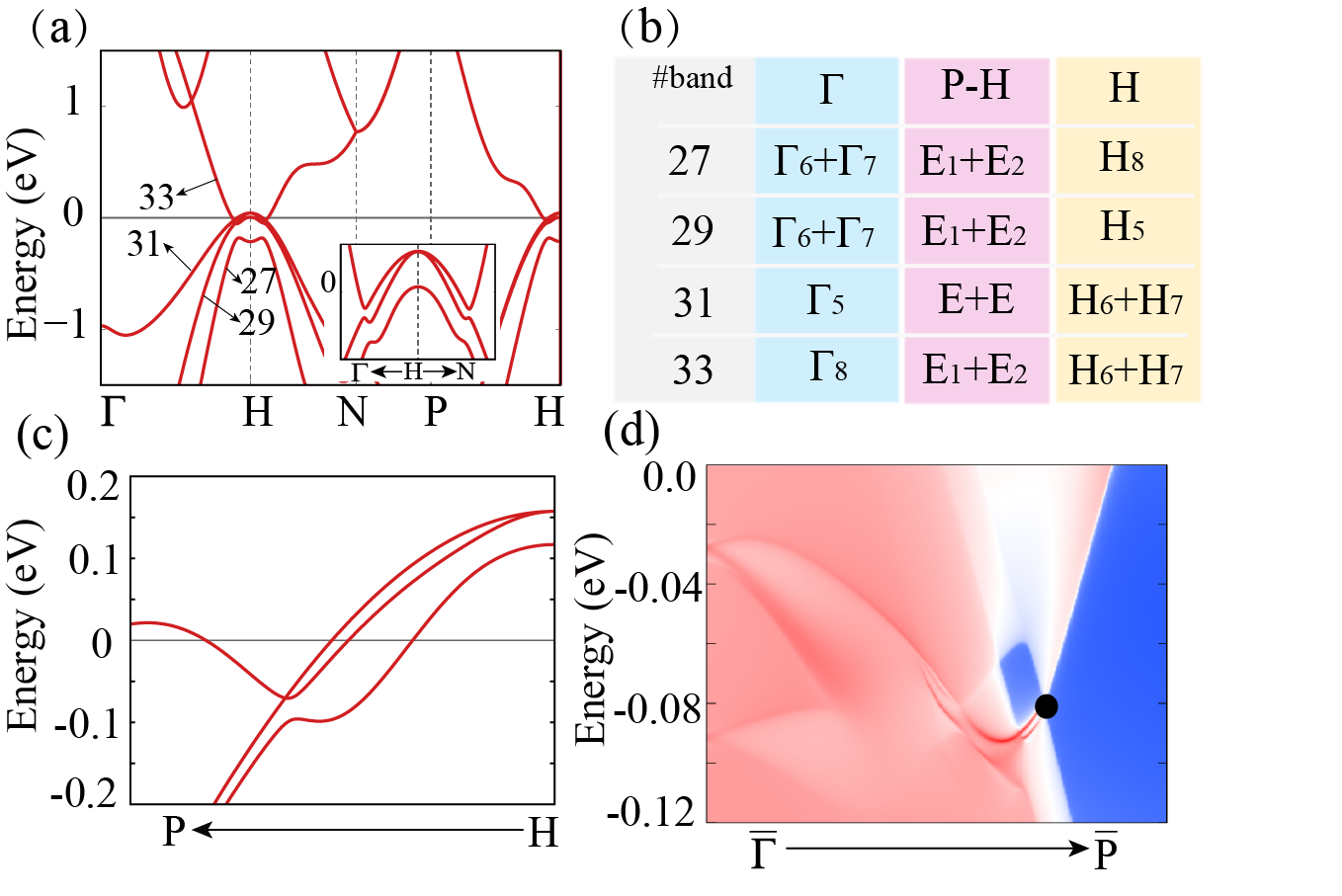}
	\caption{(a)Band structures of topological DSM.
    (b) IRRs of DSM bands around $E_F$.
    (c) Enlarged band structures along $P-H$ around the Dirac point.
    (d) The calculated LDOS of DSM on its $(001)$ surface of the unit cell,
    where black point represents location of projected Dirac point.
    \label{fig:fig4}}
\end{figure}

\textit{Dirac semimetal} ---We next discuss the DSM phase,
which can be obtained by applying a compressive strain to the topological LSM phase.
The internal coordinate $x = 0.1018$ is adopted, which has been reported in another experiment~\cite{kurakevych2016synthesis}.
The calculated band structures and corresponding IRRs
of DSM are shown in Fig.~\ref{fig:fig4}(a) and~\ref{fig:fig4}(b).
Our results indicate that the band inversion between antibonding and bonding states is enhanced as $x$ increasing,
which makes the fourfold degenerate node lift above the $E_F$.
As a result, the low energy physics is determined by the crossing between
$|J_z=\pm\frac{3}{2}\rangle^{+} $ and $| J_z=\pm\frac{1}{2}\rangle^{-}$ states.
In general, the SOC interaction would open a hybridization gap at all the crossing point of $|J_z=\pm\frac{3}{2}\rangle^{+} $ and $| J_z=\pm\frac{1}{2}\rangle^{-}$ states as shown
in the inset of Fig.~\ref{fig:fig4}(a).
However, on the $H-P$ path, due to the presence of $C_3^{111}$,
 $|J_z=\pm\frac{3}{2}\rangle^+$ states have the $E+E$~\cite{aroyo2011crystallography} IRRs,
 while $|J_z=\pm\frac{1}{2}\rangle^-$ states have $E_1 +E_2$~\cite{aroyo2011crystallography} IRRs.
 Therefore, they could cross each other exactly and form a Dirac point
 close to the $E_F$ as shown in Fig.~\ref{fig:fig4}(c).
 According to the above discussions, the DSM phase in BC8-Si
 is similar to that of $\mathrm{Na_3Bi}$,
 both of them are protected by $C_3$, $I$ and $\mathcal{T}$~\cite{wang2012dirac}.
Therefore, the BC8-Si becomes a topological DSM  with $x$ exceeding 0.1008,
and Fermi arcs formed by the topological surface states
would be expected on its surface.
By constructed the maximally localized Wannier functions,
we carry out the Green's function calculations on the semi-infinite (001) surface of the unit cell,
and plot the corresponding LDOS in Fig.~\ref{fig:fig4}(d),
which evidently shows two Fermi arcs originated from the projected Dirac point
and buried into the bulk states.
We note that the presence of two Fermi arcs in Fig.~\ref{fig:fig4}(d)
is because two bulk Dirac points are projected
to the same point on (001) surface as marked in Fig.~\ref{fig:fig4}(d).

\textit{Conclusion}---The electronic and topological properties of BC8-Si are explored by the first-principles calculations and model analysis. It demonstrates that BC8-Si is a topological LSM characterized by the band inversion at $H$ point and three quadratic nodes are located at $E_F$ exactly, which can holds the stabilized topological surface state on its (001) surface of the unit cell. Our calculations further suggest that LSM can be tuned to a NI or topological DSM by tiny changing of the crystal parameters, which can be achieved by the variation of the applied pressure during crystal synthesis. These results can well explain the previous contrary reports on the electronic properties of BC8-Si. More importantly, the topological surface states in BC8-Si could be a good connection between the topological quantum devices and silicon chips, which will stimulate more efforts on the promising electric devices.


\textit{Acknowledgments}---This work was supported by the National Key Research
and Development Program of China (2018YFA0307000),
and the National Natural Science Foundation of China (11874022).


\begin{widetext}
\section{APPENDIX A: $k \cdot p$ model}
The matrix form of the generators under the basis of $| p_x\rangle^+$, $|p_y\rangle^+$, $| p_z\rangle^+$ and $|s\rangle^-$ are given:
\begin{align} \label{eq:1}
	\begin{split}
		C_3^{111} =\begin{bmatrix}
0&0&1&0\\
1&0&0&0\\
0&1&0&0\\
0&0&0&1
\end{bmatrix}\!,
I =\begin{bmatrix}
1&0&0&0\\
0&1&0&0\\
0&0&1&0\\
0&0&0&-1
\end{bmatrix}\!,
C_{2y} =\begin{bmatrix}
1&0&0&0\\
0&-1&0&0\\
0&0&1&0\\
0&0&0&-1
\end{bmatrix}\!,
\mathcal{T} =\begin{bmatrix}
1&0&0&0\\
0&1&0&0\\
0&0&1&0\\
0&0&0&1
\end{bmatrix}K\!,
	\end{split}
\end{align}
where $K$ represents the complex conjugate operation. Hamiltonian at $H$ point without spin-orbit coupling(SOC) is constructed with the generators and method of invariants as follows((up to quadratic order in $k$):

\begin{align}
H_{non-SOC} =\begin{bmatrix}
E_b+t_3k_x^2+t_2k_y^2+t_1k_z^2&t_6k_xk_y&t_6k_xk_z&it_5k_x\\
t_4k_xk_y&E_b+t_1k_x^2+t_3k_y^2+t_2k_z^2&t_4k_yk_z&it_5k_y\\
t_6k_xk_z&t_6k_yk_z&E_b+t_2k_x^2+t_1k_y^2+t_3k_z^2&it_5k_z\\
-it_5k_x&-it_5k_y&-it_5k_z&E_{ab}+t_4(k_x^2+k_y^2+k_z^2)
\end{bmatrix}\!.
\label{eq:H_{non-SOC}}
\end{align}
The parameters of model in Eq.~\ref{eq:H_{non-SOC}} can be obtained by fitting band structures from first-principles calculations.
First of all, we will consider
the band structures along $k_x$ direction. The non-SOC model along $k_x$ direction, i.e. $k_y=k_z=0$:
\begin{align}
H_{non-SOC}(k_x) =\begin{bmatrix}
E_b+t_3k_x^2&0&0&it_5k_x\\
0&E_b+t_1k_x^2&0&0\\
0&0&E_b+t_2k_x^2&0\\
-it_5k_x&0&0&E_{ab}+t_4k_x^2
\end{bmatrix}\!,
\label{eq:H_{non-SOCx}}
\end{align}

with the value of $t_1, t_2, t_3, t_4$ and $t_5$ given in Table.I, the energy eigenvalues of four-band model are well fitted to the band structures from first-principles calculations as shown in Fig.~\ref{fig:non-SOCfit}(a).

To obtained the value of $t_6$, the band structures along $k_x=k_y, k_z=0$ directions are considered. The corresponding model reads:
\begin{align}
H_{non-SOC}(k_xk_y) =\begin{bmatrix}
E_b+t_3k_x^2+t_2k_y^2&t_6k_xk_y&0&it_5k_x\\
t_6k_xk_y&E_b+t_1k_x^2+t_3k_y^2&0&it_5k_y\\
0&0&E_b+t_2k_x^2+t_1k_y^2&0\\
-it_5k_x&-it_5k_y&0&E_{ab}+t_4(k_x^2+k_y^2)
\end{bmatrix}\!.
\label{eq:H_{non-SOCy}}
\end{align}
With the values of $t_6$ given in Table.I, the energy eigenvalues of four-band model are well fitted to the band structures from first-principles calculations as shown in Fig.~\ref{fig:non-SOCfit}(b).

Finally, band structures along $k_x=k_y=k_z$ and high symmetry lines are considered to check the model parameters that derived from $k_x$ and $k_x=k_y$ lines as shown in Fig.~\ref{fig:non-SOCfit}(c) and (d).

\begin{figure}
\includegraphics[width=0.7\columnwidth]{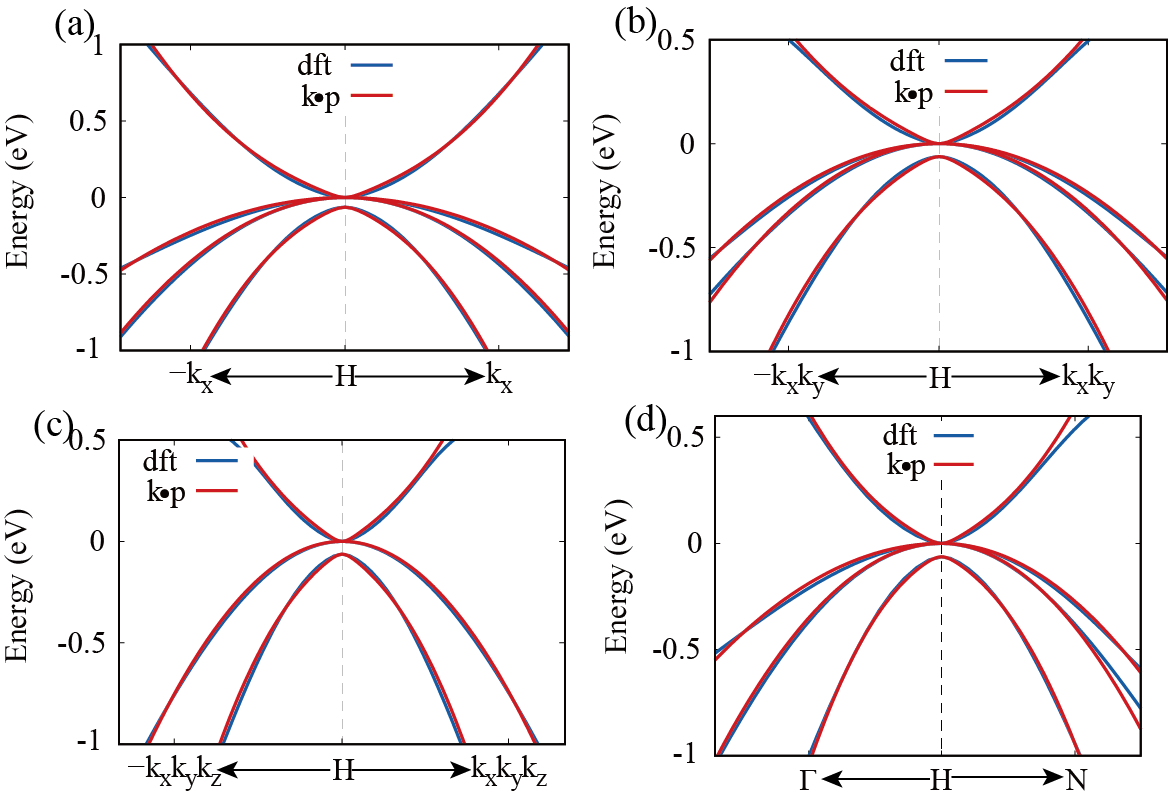}
\caption{The fitted band structures of $k \cdot p$ and first-principles calculations without SOC.
\label{fig:non-SOCfit}}
\end{figure}

Considering SOC, only the atomic SOC strength parameter $\lambda$ is counted in Eq.\ref{eq:p_{soc}}. The SOC effect under the basis of $\phi$ [ $| p_x\rangle^+$, $|p_y\rangle^+$, $| p_z\rangle^+$, $|s\rangle^-$) $\otimes$ ($|\uparrow\rangle$  $|\downarrow\rangle$ ] is written:

\begin{align}
H_{soc} =\lambda\begin{bmatrix}
0&-i&0&0&0&0&1&0\\
i&0&0&0&0&0&-i&0\\
0&0&0&0&-1&i&0&0\\
0&0&0&0&0&0&0&0\\
0&0&-1&0&0&i&0&0\\
0&0&-i&0&-i&0&0&0\\
1&i&0&0&0&0&0&0\\
0&0&0&0&0&0&0&0
\end{bmatrix}\!.
\label{eq:p_{soc}}
\end{align}

The eight-band Hamiltonian with SOC is obtained in Eq.~\ref{eq:Hsoc}.
\begin{equation}
H=\sigma_{0}\otimes H_{non-SOC}+H_{soc}
\label{eq:Hsoc}
\end{equation}

With $\lambda$=0.0152~eV, the energy eigenvalues of eight-band model are well fitted to the band structures from first-principles calculations as shown in Fig.~\ref{fig:socfit},  which means  the low energy physics around $H$ point can be described well by our model and parameters in Table.I.

\begin{figure}
\includegraphics[width=0.7\columnwidth]{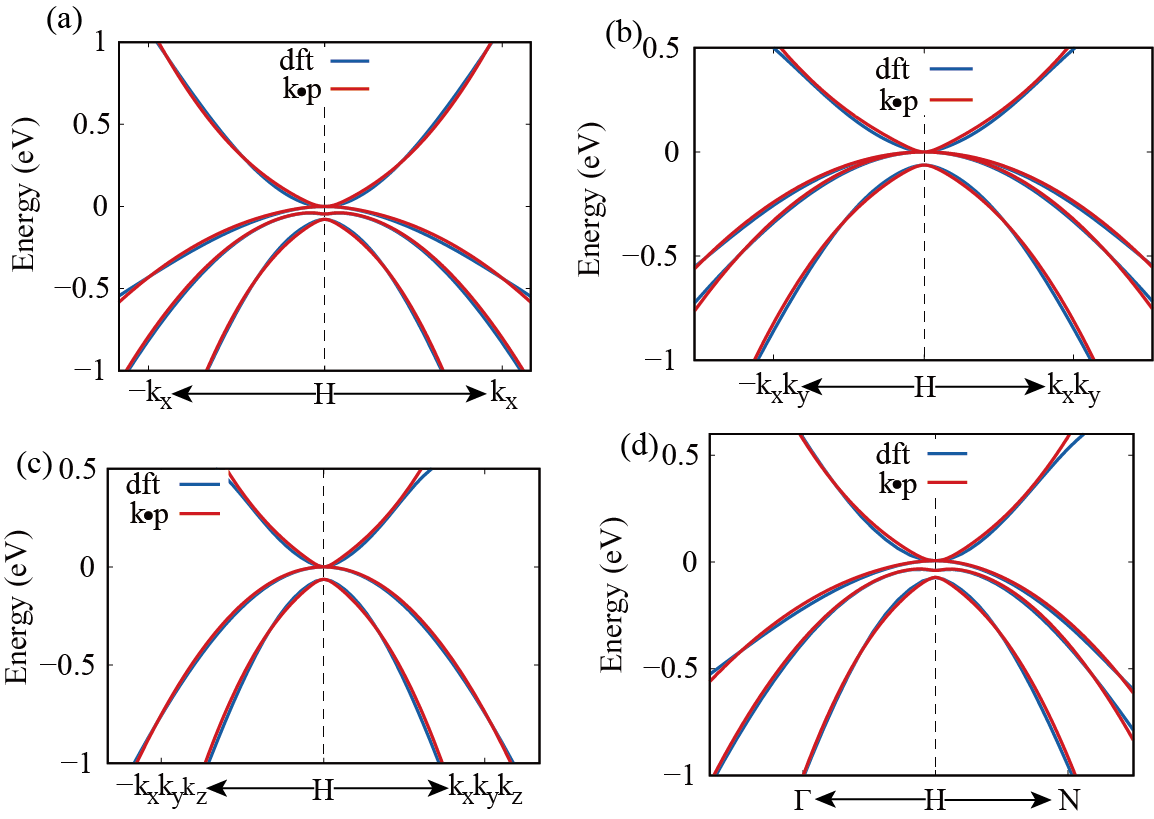}
\caption{The fitted band structures of $k \cdot p$ and first-principles calculations with SOC.
\label{fig:socfit}}
\end{figure}

\section{APPENDIX B:  Basis transformation}
Considering SOC, the eigenstates can be written as $| J, J_z\rangle$ representation, where $J$, $J_z$ indicate total angular momentum and its z-direction component.  In order to transform the basis $\phi$ to $\psi$ [ $| \frac{3}{2},  \frac{1}{2}\rangle^{+} $, $| \frac{3}{2},  \frac{3}{2}\rangle^{+} $, $| \frac{3}{2},  -\frac{3}{2}\rangle^{+} $, $| \frac{3}{2}, - \frac{1}{2}\rangle^{+} $, $| \frac{1}{2},  \frac{1}{2}\rangle^{+} $, $| \frac{1}{2},  -\frac{1}{2}\rangle^{+} $, $| \frac{1}{2},  \frac{1}{2}\rangle^{-} $ and $| \frac{1}{2}, - \frac{1}{2}\rangle^{-} $ ], the specific combination form between two bases are as follows:

		\begin{equation} \label{eq:u}
			\begin{aligned}
				| \frac{1}{2},\frac{1}{2}\rangle =& | is,\uparrow\rangle ,\\
                     | \frac{1}{2}, - \frac{1}{2}\rangle =& | is,\downarrow\rangle,\\
                     | \frac{3}{2}, \frac{3}{2}\rangle =&\frac{1}{\sqrt2} | -(p_x+ip_y),\uparrow\rangle,\\
                     | \frac{3}{2}, -\frac{3}{2}\rangle =&\frac{1}{\sqrt2} | (p_x-ip_y),\downarrow\rangle,\\
                     | \frac{3}{2}, \frac{1}{2}\rangle =&\frac{1}{\sqrt6} |- (p_x+ip_y),\downarrow\rangle+\sqrt{\frac{2}{3}}|p_z,\uparrow\rangle,\\
                     | \frac{3}{2}, -\frac{1}{2}\rangle =&\frac{1}{\sqrt6} | (p_x-ip_y),\uparrow\rangle+\sqrt{\frac{2}{3}}|p_z,\downarrow\rangle,\\
                     | \frac{1}{2}, \frac{1}{2}\rangle =&\frac{1}{\sqrt3} |- (p_x+ip_y),\downarrow\rangle-\frac{1}{\sqrt{3}}|p_z,\uparrow\rangle,\\
                     | \frac{1}{2}, -\frac{1}{2}\rangle =&-\frac{1}{\sqrt3} |(p_x-ip_y),\uparrow\rangle+\frac{1}{\sqrt{3}}|p_z,\downarrow\rangle.
			\end{aligned}
		\end{equation}

According to Eq.~\ref{eq:u},the transformation matrix $u$ reads:

\begin{align}
u =\begin{bmatrix}
0&-\frac{1}{\sqrt{2}}&0&\frac{1}{\sqrt{6}}&-\frac{1}{\sqrt{3}}&0&0&0\\
0&-\frac{i}{\sqrt{2}}&0&-\frac{i}{\sqrt{6}}&\frac{i}{\sqrt{3}}&0&0&0\\
\sqrt{\frac{2}{3}}&0&0&0&0&-\frac{1}{\sqrt{3}}&0&0\\
0&0&0&0&0&0&i&0\\
-\frac{1}{\sqrt{6}}&0&\frac{1}{\sqrt{2}}&0&0&-\frac{1}{\sqrt{3}}&0&0\\
-\frac{i}{\sqrt{6}}&0&-\frac{i}{\sqrt{2}}&0&0&-\frac{i}{\sqrt{3}}&0&0\\
0&0&0&\sqrt{\frac{2}{3}}&\frac{1}{\sqrt{3}}&0&0&0\\
0&0&0&0&0&0&0&i
\end{bmatrix}\!,
\label{eq:U}
\end{align}

The trace of Hamiltonian is same, $uHu^{-1}=H^\prime$, we get
\begin{align}
H^\prime =\begin{bmatrix}
M_0 (k) & -\frac{t_6 k_+ kz}{\sqrt{3}} & V_1 (k)& 0 & -\frac{t_6 k_- kz}{\sqrt{2}} & V_2 (k)  &-\sqrt{\frac{2}{3}}t_5 k_z & \frac{t_5 k_- }{\sqrt{6}}  \\
				-\frac{t_6 k_- kz}{\sqrt{3}} & M_1 (k) & 0 & V_1 (k) & V_3 (k) & \frac{t_6 k_- kz}{\sqrt{6}} & \frac{t_5 k_- }{\sqrt{2}} & 0 \\
				V_1^{*}(k) & 0 & M_1 (k) & \frac{t_6 k_+ kz}{\sqrt{3}} & \frac{t_6 k_+ kz}{\sqrt{6}} & -V_3^* (k) & 0 & -\frac{t_5 k_+ }{\sqrt{2}} \\
				0	 & V_1^{*}(k) & \frac{t_6 k_- kz}{\sqrt{3}} & M_0 (k) & -V_2 (k) & -\frac{t_6 k_+ kz}{\sqrt{2}} &  -\frac{t_5 k_+ }{\sqrt{6}}  & -\sqrt{\frac{2}{3}}t_5 k_z \\
				-\frac{t_6 k_+ kz}{\sqrt{2}} & V_3^{*}(k) & \frac{t_6 k_- kz}{\sqrt{6}} & -V_2^{*}(k) & M_2 (k) & 0 & \frac{t_5 k_+ }{\sqrt{3}} & -\frac{t_5 k_z}{\sqrt{3}}    \\
				V_2^{*}(k) & \frac{t_6 k_+ kz}{\sqrt{6}} & -V_3(k) & -\frac{t_6 k_- kz}{\sqrt{2}} & 0 & M_2 (k) & \frac{t_5 k_z}{\sqrt{3}}  & \frac{t_5 k_- }{\sqrt{3}} \\
				-\sqrt{\frac{2}{3}}t_5 k_z & \frac{t_5 k_+ }{\sqrt{2}} & 0 &-\frac{t_5 k_- }{\sqrt{6}} & \frac{t_5 k_- }{\sqrt{3}} & \frac{t_5 k_z}{\sqrt{3}} & M_3 (k) &  0\\
				\frac{t_5 k_+ }{\sqrt{6}} & 0 &  -\frac{t_5 k_- }{\sqrt{2}}  & -\sqrt{\frac{2}{3}}t_5 k_z   & -\frac{t_5 k_z}{\sqrt{3}} & \frac{t_5 k_+ }{\sqrt{3}}  & 0 & M_3 (k) \\
\end{bmatrix}\!,
\label{eq:H_{basis8}}
\end{align}
		where $k_+ = k_x + ik_y $, $k_- = k_x - ik_y $ and
		\begin{equation} \label{eq:appendix}
			\begin{aligned}
				& M_0(k)=\frac{6E_b+6\lambda + (t_1 +4t_2 +t_3 ) k^2_{x} + (4t_1 +t_2 +t_3 )k^2_{y} + (t_1 +t_2 +4t_3 )k^2_{z}}{6} \\
				& M_1 (k)=\frac{2E_b+2\lambda + t_1(k^2_{x} +k^2_{z} ) + t_2(k^2_{y} +k^2_{z} ) + t_3(k^2_{x} +k^2_{y} )}{2} \\
				& M_2 (k)=\frac{3E_b - 6\lambda + (t_1 + t_2 + t_3 )(k^2_{x} +k^2_{y} +k^2_{z} )}{3} \\
				& M_3 (k)=E_{ab} + t_4 (k^2_{x} +k^2_{y} +k^2_{z} ) \\
                     & V_1 (k)=\frac{k^{2}_{z} (-t_1 + t_2 ) + k^{2}_{y} (-t_2 + t_3 ) + k^{2}_{x} (t_1 - t_3 ) + 2it_6 k_x k_y }{2\sqrt{3}} \\
				& V_2 (k)=\frac{k^{2}_{z} (t_1 + t_2 - 2t_3 ) + k^{2}_{x} (t_1 - 2t_2 + t_3 ) + k^{2}_{y} (-2t_1 + t_2 + t_3 )}{3\sqrt{2}} \\
				& V_3 (k)=\frac{k^{2}_{z} (t_1 - t_2 ) + k^{2}_{y} (t_2 - t_3 ) + k^{2}_{x} (-t_1 + t_3 ) - 2it_6 k_x k_y }{\sqrt{6}}.
			\end{aligned}
		\end{equation}

From Eq.~\ref{eq:H_{basis8}} and ~\ref{eq:appendix}, the onset energy difference between $| J=\frac{3}{2}\rangle^+$ and $| J=\frac{1}{2}\rangle^+ $ is $3\lambda$. Therefore, the value of $\lambda$ is determined by the on site energy of the band structures from first-principles calculations, i.e. $\lambda$=0.0152~eV which is consistent with the values in Table.I.

\section{APPENDIX C: Downfolding}
In order to explore the topological properties, the eight-band model in Eq.~\ref{eq:H_{basis8}} is downfolded to four-band Hamiltonian under the basis of $| \frac{3}{2},  \frac{1}{2}\rangle^{+} $, $| \frac{3}{2},  \frac{3}{2}\rangle^{+} $, $| \frac{3}{2},  -\frac{3}{2}\rangle^{+} $ and $| \frac{3}{2}, - \frac{1}{2}\rangle^{+} $ up to the leading term of $k$.
The influences from $|J= \frac{1}{2}\rangle^{-}$ orbitals to $|J= \frac{3}{2}\rangle^+ $ subspace are second-order of $k$ terms while the influences from $|J= \frac{1}{2}\rangle^{+}$ orbitals are fourth-order of $k$ terms. Therefore, the influences from $| J=\frac{1}{2}\rangle^{+}$ orbitals are ignored while only the influences from $|J= \frac{1}{2}\rangle^{-}$ orbitals are considered.
By perturbation theory,
\begin{equation}
H_{4\times4}= H_{ul} (k) + H_{\frac{3}{2} \frac{1}{2}}(p) \frac{1}{H_{\frac{3}{2}\frac{3}{2}}(0)-H_{\frac{1}{2}\frac{1}{2}}(0)}H_{\frac{1}{2} \frac{3}{2}}(p)
\label{eq:H_Lo}\!.
\end{equation}
Among them, $ H_{ul} (k) $ comes from the upper left $4 \times 4$ block of the eight-band model in Eq.~\ref{eq:H_{basis8}}. $H_{\frac{1}{2} \frac{3}{2}}(p)$, $H_{\frac{3}{2} \frac{1}{2}}(p)$ and $\frac{1}{H_{\frac{3}{2}\frac{3}{2}}(0)-H_{\frac{1}{2}\frac{1}{2}}(0)}$ are writeen

\begin{align}
H_{\frac{1}{2} \frac{3}{2}}(p) =\begin{bmatrix}
 -\sqrt{\frac{2}{3}}t_5 k_z & \frac{t_5 k_+ }{\sqrt{2}} & 0 &-\frac{t_5 k_- }{\sqrt{6}}\\
\frac{t_5 k_+ }{\sqrt{6}} & 0 &  -\frac{t_5 k_- }{\sqrt{2}}  & -\sqrt{\frac{2}{3}}t_5 k_z
\end{bmatrix}\!,
\label{eq:$H_op}
\end{align}

\begin{align}
H_{\frac{3}{2} \frac{1}{2}}(p)=\begin{bmatrix}
-\sqrt{\frac{2}{3}}t_5 k_z & \frac{t_5 k_- }{\sqrt{6}} \\
\frac{t_5 k_- }{\sqrt{2}} & 0 \\
 0 & -\frac{t_5 k_+ }{\sqrt{2}} \\
 -\frac{t_5 k_+ }{\sqrt{6}}  & -\sqrt{\frac{2}{3}}t_5 k_z
\end{bmatrix}\!,
\label{eq:H_2}
\end{align}

\begin{align}
 \frac{1}{H_{\frac{3}{2}\frac{3}{2}}(0)-H_{\frac{1}{2}\frac{1}{2}}(0)}=\begin{bmatrix}
\frac{1}{-\Delta}&0\\
0&\frac{1}{-\Delta}
\end{bmatrix}\!,
\label{eq:H_2}
\end{align}
where $\Delta=E_{ab}-E_b-\lambda$ is the energy difference between $|J=\frac{1}{2}\rangle^-$ and $|J= \frac{3}{2}\rangle^+ $ states.

The second term in Eq.~\ref{eq:H_Lo}, i.e. the perturbation term from $|J=\frac{1}{2}\rangle^-$ orbitals is
\begin{align}
H_c(p)=\frac{t_5^2}{-\Delta}\begin{bmatrix}
\frac{1}{6}k_+k_- +\frac{2k_z^2}{3} & -\frac{k_+k_z}{\sqrt{3}} & -\frac{k_-^2}{2\sqrt{3}} & 0\\
-\frac{k_- kz}{\sqrt{3}} & \frac{k_+k_-}{2} & 0 & -\frac{k_-^2}{2\sqrt{3}} \\
-\frac{k_+^2}{2\sqrt{3}}	 & 0 & \frac{k_-k_+}{2}&  \frac{k_+k_z}{\sqrt{3}}\\
0	 & -\frac{k_+^2}{2\sqrt{3}}  & \frac{k_-k_z}{\sqrt{3}} & \frac{1}{6}(k_-k_+) +\frac{2k_z^2}{3}
\end{bmatrix}\!,
\label{eq:H_c}
\end{align}

Finally, the four-band Hamiltonian of BC8-Si under the basis of $| \frac{3}{2},  \frac{1}{2}\rangle^{+} $, $| \frac{3}{2},  \frac{3}{2}\rangle^{+} $, $| \frac{3}{2},  -\frac{3}{2}\rangle^{+} $ and $| \frac{3}{2}, - \frac{1}{2}\rangle^{+} $ is written as
\begin{align}
H_{4\times4} =\begin{bmatrix}
M_0 (k) & -\frac{t_6 k_+ kz}{\sqrt{3}} & V_1 (k)& 0\\
-\frac{t_6 k_- kz}{\sqrt{3}} & M_1 (k) & 0 & V_1 (k) \\
V_1^{*}(k) & 0 & M_1 (k) & \frac{t_6 k_+ kz}{\sqrt{3}} \\
0	 & V_1^{*}(k) & \frac{t_6 k_- kz}{\sqrt{3}} & M_0 (k)
\end{bmatrix}\! +\frac{t_5^2}{-\Delta}\begin{bmatrix}
\frac{1}{6}k_+k_- +\frac{2k_z^2}{3} & -\frac{k_+k_z}{\sqrt{3}} & -\frac{k_-^2}{2\sqrt{3}} & 0\\
-\frac{k_- kz}{\sqrt{3}} & \frac{k_+k_-}{2} & 0 & -\frac{k_-^2}{2\sqrt{3}} \\
-\frac{k_+^2}{2\sqrt{3}}	 & 0 & \frac{k_-k_+}{2}&  \frac{k_+k_z}{\sqrt{3}}\\
0	 & -\frac{k_+^2}{2\sqrt{3}}  & \frac{k_-k_z}{\sqrt{3}} & \frac{1}{6}(k_-k_+) +\frac{2k_z^2}{3}
\end{bmatrix}\!
\label{eq:H_downfolding}
\end{align}

The downfolding model in Eq.~\ref{eq:H_downfolding} can be written as $\Gamma$ matrix.
\begin{equation}\label{downfolding-gamma}
H_{4\times4}= H_{ul} (k)+\frac{t_5^2}{-\Delta}(\frac{k_-k_+ + k_z^2}{3}\Gamma_0
+(-\frac{k_-k_+}{6} + \frac{k_z^2}{3})\Gamma_1 - \frac{k_xk_z}{\sqrt{3}}\Gamma_2
+ \frac{k_yk_z}{\sqrt{3}}\Gamma_3 - \frac{k_x^2-k_y^2}{2\sqrt{3}}\Gamma_4 - \frac{2k_xk_y}{2\sqrt{3}}\Gamma_5)
\end{equation}
The $\Gamma$ matrixes are defined as 
$\Gamma_0 =\tau_0\otimes\sigma_0$,  $\Gamma_1 =\tau_z\otimes\sigma_z$, $\Gamma_2 =\tau_z\otimes\sigma_x$, $\Gamma_3 =\tau_z\otimes\sigma_y$, $\Gamma_4 =\tau_x\otimes\sigma_0$, $\Gamma_5 =\tau_y\otimes\sigma_0$.  $\tau_{0}$ and $\sigma_{0}$ are the $ 2 \times 2 $ identity matrices, $\tau_{x,y,z}$ and $\sigma_{x,y,z}$ reprent the orbital and spin space, respectively.

\section{APPENDIX D: Topological characters}
Maxim Kharitonov et al. have
confirmed that the four-band 3D LSM are topological, if its 2D reductions to some
planes in momentum space passing the quadratic node
are topologically nontrivial~\cite{kharitonov2017universality}.
The topological property of the reduced 2D nodal semimetal is determined by
the phase diagram in the ($\lvert\frac{\beta_0}{\beta\perp}\rvert$, $\lvert\frac{\beta_z}{\beta\perp}\rvert$) parameter plane, the 2D model in ref~\cite{kharitonov2017universality} is
\begin{align}
H_2^\beta(p_x, p_y) =\begin{bmatrix}
(\beta_0 + \beta_z)p_+p_- &\beta_\perp p_-^2\\
\beta_\perp p_+^2&(\beta_0 - \beta_z)p_+p_-
\end{bmatrix}\!.
\label{eq:H_{PRL}}
\end{align}
Next, we show the estimation of ${\beta_0}$, ${\beta_z}$ and ${\beta_\perp}$ in BC8-Si.
In $k_xk_y$ plane, the $4\times4$ Hamiltonian in Eq.~\ref{eq:H_downfolding} are decoupled into two pairs due to the $\mathcal{T}$ symmetry. The parameters ${\beta_0}$, ${\beta_z}$ and ${\beta_\perp}$ calculated from any one of the two $2\times2$ Hamiltonian are same as the parameters calculated from the other one.
For example, $2\times2$ Hamiltonian under the basis of $| \frac{3}{2},  \frac{3}{2}\rangle^{+}|$, $| \frac{3}{2},  -\frac{1}{2}\rangle^{+} $ is considered
\begin{align}
H_{2\times2} =\begin{bmatrix}
\frac{2E_b+2\lambda + (t_1+t_3)k^2_{x}  + (t_2+t_3)k^2_{y}}{2}  & \frac{k^{2}_{y} (-t_2 + t_3 ) + k^{2}_{x} (t_1 - t_3 ) + 2it_6 k_x k_y }{2\sqrt{3}} \\
\frac{ k^{2}_{y} (-t_2 + t_3 ) + k^{2}_{x} (t_1 - t_3 ) - 2it_6 k_x k_y }{2\sqrt{3}}  & \frac{6E_b+6\lambda + (t_1 +4t_2 +t_3 ) k^2_{x} + (4t_1 +t_2 +t_3 )k^2_{y}}{6}
\end{bmatrix}\! +\frac{t_5^2}{-\Delta}\begin{bmatrix}
\frac{k_+k_-}{2}& -\frac{k_-^2}{2\sqrt{3}} \\
-\frac{k_+^2}{2\sqrt{3}}  & \frac{1}{6}(k_-k_+) 
\end{bmatrix}\!.
\label{eq:H_N2}
\end{align}
Here, the first term describes the electronic anisotropy of the crystal. However, the second term takes the leading role because of the small $\Delta$. Moreover, one can further rewrite Eq. \ref{eq:H_N2} as the following form 
\begin{align}
H_{2\times2} =\begin{bmatrix}
\frac{(t_1+t_3)+(t_2+t_3)}{4}k_+k_-& \frac{(-t_2 + t_3 )+(t_1 - t_3 )}{4\sqrt{3}}k_-^2 \\
 \frac{(-t_2 + t_3 )+(t_1 - t_3 )}{4\sqrt{3}}k_+^2  & \frac{ (t_1 +4t_2 +t_3 ) + (4t_1 +t_2 +t_3 )}{12}k_+k_-
\end{bmatrix}\! +\frac{t_5^2}{-\Delta}\begin{bmatrix}
\frac{k_+k_-}{2}& -\frac{k_-^2}{2\sqrt{3}} \\
-\frac{k_+^2}{2\sqrt{3}}  & \frac{1}{6}(k_-k_+) 
\end{bmatrix}\!+H_{anis},
\label{eq:H_N3}
\end{align}
where, the first two terms are both isotropic as Eq.~\ref{eq:H_{PRL}}, and the anisotropic effect is attributed into $H_{anis}$. Using the isotropic coefficients in Eq.~\ref{eq:H_N3} and the parameters in Table I of the main text, one can estimate $\beta_0= 13.7854$, $\beta_z=10.1978$,  ${\beta_\perp}=-17.6633$, and  $\frac{|\beta_0|}{|\beta_\perp|}=0.7804, \frac{|\beta_z|} {|\beta_\perp|}$=0.5733, which strongly imply that BC8-Si is a topological LSM, similar as HgTe and $\alpha$-Sn.
\end{widetext}
\bibliography{refs}

\end{document}